\documentstyle[aps,epsf]{revtex}
\begin{document}
\draft

\wideabs{
\title{
Thermoelastic noise and homogeneous thermal noise in finite sized 
gravitational-wave test masses
}
\author{Yuk Tung Liu and Kip S.\ Thorne
}
\address{
Theoretical Astrophysics, California Institute of Technology, Pasadena, CA 91125
}
\date{Submitted to Physical Review D15 on February 15, 2000}
\maketitle
\begin{abstract}

An analysis is given of thermoelastic noise (thermal noise due to 
thermoelastic dissipation) in finite sized test masses of laser interferometer 
gravitational-wave detectors.  Finite-size
effects increase the thermoelastic
noise by a modest amount; for example, for the sapphire test masses 
tentatively planned for LIGO-II and plausible beam-spot radii,
the increase is $\alt 10$ per cent.  As a side issue, errors are pointed
out in the currently used formulas for conventional, homogeneous
thermal noise (noise associated with dissipation which is homogeneous and 
described by an imaginary part
of the Young's modulus) in finite sized test
masses. Correction of these errors increases the 
homogeneous thermal noise by $\alt 5$ per cent for LIGO-II-type configurations.

\end{abstract}
}
\narrowtext
%\twocolumn

\section{INTRODUCTION AND SUMMARY}
\label{sec:Intro}

Braginsky, Gorodetsky and Vyatchanin \cite{BGV} (henceforth BGV)
have recently shown
that noise associated with thermoelastic dissipation in
test masses will affect significantly the performance of LIGO-II
gravitational-wave detectors.\footnote{``LIGO-II'' refers to  the 
second generation of detectors in the Laser Interferometer Gravitational 
Wave Observatory\cite{ligo} and in its international partners.}  

The BGV computation of thermoelastic noise was based on an idealization
in which each test mass has an arbitrarily large radius and length compared to
the size of the light's beam spot on the mirrored test-mass face.  In this
limiting case, BGV showed that the spectral density $S_h(f)$
of the thermoelastic gravitational-wave noise scales as the inverse cube of 
the beam-spot radius $r_o$, $S_h \propto 1/r_o^3$,
so it is desirable to make $r_o$ large.
However, when $r_o$ is no longer small compared to the test-mass size, 
the BGV analysis breaks down.  

The principal purpose of this paper is to
explore, quantitatively, the sign and magnitude of that breakdown.
As we shall see, that breakdown (i.e., finite size of the test masses)
{\it increases} the thermoelastic noise; but for expected beam-spot radii 
($r_0\alt 3/10$ the test-mass radius $a$), the increase is modest 
($\alt 10$ per cent). 

A second purpose of this paper is to show how the BGV analysis of thermoelastic
noise can be simplified considerably; and (adapting techniques due to 
Bondu, Hello and Vinet \cite{bondu} -- henceforth BHV), to show how to 
generalize the BGV analysis to finite sized test masses.

A third purpose is to point out and correct errors in the BHV analysis of
conventional homogeneous thermal noise 
(noise associated with a small, position-independent, imaginary part of the
Young's modulus) in finite sized test masses. 
The corrections of the BHV formulae
increase homogeneous thermal noise by $\alt 5$ per cent
for beam-spot radii $\alt 3/10$ the test-mass radius $a$, and thus are
primarily of conceptual importance, not practical importance.  

In Sec.\ \ref{sec:Method}, we outline our method of computing thermoelastic
noise, in Sec.\ \ref{sec:InfTM} we use our method to verify the BGV result
for thermoelastic noise in the limit of arbitrarily large test masses,
in Sec.\ \ref{sec:FiniteTestMasses} we compute the
thermoelastic noise in finite sized test masses and estimate the
accuracy of our analysis, in Sec.\ \ref{sec:Conv}
we correct the errors in the BHV computation of conventional, 
homogeneous thermal noise, and in Sec.\ \ref{sec:Conclusion} we make some
concluding remarks. 

\section{Method of Calculation}
\label{sec:Method}

Our analysis of thermoelastic noise is a simplification of one of the
procedures developed by BGV: Appendix C of Ref.\ 
\cite{BGV}.  The foundation of 
the analysis is Levin's \cite{levin} ``direct'' method of computing 
thermal noise (of which thermoelastic noise is a special case):

Levin begins by noting that the gravitational-wave detector's laser beam
reads out a difference of generalized positions $q(t)$ of the detector's four
test masses, with each $q$ given by an average, over the beam spot's Gaussian
power profile, of the normal displacement $\delta z \equiv u_z$ 
of the test-mass face:
\begin{eqnarray}
q &=& \int_0^a\int_0^{2\pi}
{e^{-r^2/r_o^2} 
\over \pi r_o^2 \left(1-e^{-a^2/r_o^2}\right)} \delta z(r,\phi) 
\; rd\phi dr \nonumber \\
&\simeq& \int_0^a\int_0^{2\pi}
{e^{-r^2/r_o^2} 
\over \pi r_o^2 } \delta z(r,\phi) 
rd\phi dr \;. 
\label{qdef}
\end{eqnarray}  
Here $(r,\phi)$ are circular polar coordinates centered on the beam-spot
center (which we presume to be at the center of the circular test-mass face),
$r_o$ is the radius at which the spot's power flux has dropped to
$1/e$ of its central value, and $a$ is the test-mass radius.  
(The factor $e^{-a^2/r_o^2}$ must be $\ll 1$ in order to keep
diffraction losses small, so we shall approximate $1- e^{-a^2/r_o^2}$ 
by unity throughout this paper.)
Levin then appeals to a very general formulation
of the fluctuation-dissipation theorem, due to Callan and Welton
\cite{callenwelton}, to show that the test-mass thermal noise can be computed
by the following thought experiment:

We imagine applying a sinusoidally
oscillating pressure,
\begin{equation}
P = F_o {e^{-r^2/r_o^2}\over \pi r_o^2 } 
\cos(\omega t)
\label{pressure}
\end{equation}
to one face of the test mass. 
Here $F_o$ is a constant force amplitude, 
$\omega = 2\pi f$ is the angular
frequency at which one wants to know the spectral
density of thermal noise, and  
the pressure distribution (\ref{pressure}) 
has precisely the same spatial profile as that of the generalized coordinate 
$q$, whose thermal noise $S_q(f)$ one wishes to compute.

The oscillating pressure $P$ feeds energy into the test mass, where it
gets dissipated by thermoelastic heat flow.
We compute the rate 
of this energy dissipation, $W_{\rm diss}$, averaged
over the period $2\pi/\omega$ of the pressure 
oscillations.\footnote{It is here that our analysis
is simpler than that of BGV. Instead of computing $W_{\rm diss}$ and using
Eq.\ (\ref{SqFD}) for the thermal noise, BGV compute the imaginary part 
$\Im(\chi)$ of the
test-mass susceptibility $\chi$
(which is much harder to compute than $W_{\rm diss}$)
and then evalute $S_q$ in terms of $\Im(\chi)$ [their Eq.\ (14)].}  
Then the fluctuation-dissipation
theorem states 
that the spectral density of the noise $S_q(f)$ is given by
\begin{equation}
S_q(f) = {8 k_{\rm B}T W_{\rm diss} \over F_o^2 \omega^2}\;
\label{SqFD}
\end{equation}
(Eq.\ (2) of Ref.\ \cite{levin}); here $k_{\rm B}$ is Boltzman's constant 
The interferometer's gravitational-wave signal $h(t)$ is the difference of the
generalized positions $q$ of the four test masses, divided by the
interferometer arm length $L$.  Correspondingly the contribution
of the test-mass thermoelastic noise to the gravitational-wave noise is
$1/L^2$ times the sum of $S_q(f)$ over the four test masses (which might have
different beam-spot sizes and thus different noises): 
\begin{equation}
S_h(f) = \sum_{A=1}^4 {S_{q_A}(f) \over L^2} \;. 
\label{ShFD}
\end{equation}

The rate $W_{\rm diss}$ of thermoelastic dissipation is given by the
following standard expression (first term of Eq.\ (35.1) of Landau and
Lifshitz \cite{LL}, cited henceforth as LL):
\begin{equation}
W_{\rm diss} = \left\langle {T dS\over dt}\right\rangle  = \left\langle
\int {\kappa \over T} (\vec \nabla \delta T)^2
\; r d\phi dr dz\right\rangle\;.
\label{WdissFD}
\end{equation}
Here the integral is over the entire test-mass interior using cylindrical
coordinates;
$T$ is the unperturbed temperature of the test-mass material and $\delta
T$ is the temperature perturbation produced by the oscillating pressure; 
$dS/dt$ is the rate of 
increase of the test mass's entropy due to the flux of heat $-\kappa \vec
\nabla \delta T$ flowing down the  
temperature gradient $\vec \nabla \delta T$, $\kappa$ is the material's 
coefficient of thermal conductivity, and $\langle ... \rangle$ denotes an
average over the pressure's oscillation period $1/f = 2\pi/\omega$.
(For conceptual clarity we explicitly write the average 
$\langle ... \rangle$ throughout this paper, even though in practice
it gives just a simple factor $\langle \cos^2 \omega t \rangle = 1/2$.)

To compute the thermal noise, then, we must calculate the 
oscillating temperature perturbation $\delta T(r,\phi,z,t)$
inside the test mass, perform the integral
(\ref{WdissFD}) over the test-mass interior and the time average
to obtain the dissipation rate, then plug that rate into Eq.\  
(\ref{SqFD}) and then Eq.\ (\ref{ShFD}).

The computation of the oscillating temperature perturbation is made
fairly simple by two well-justified approximations \cite{BGV}: 

{\it First:} The radius 
and length of the test mass are $a \sim H \sim 14$ cm and the speeds of
sound in the test-mass material are $c_s\sim 5$ km/s, so the time required for
sound to travel across the test mass is $\tau_{\rm sound} \sim 30$ $\mu$s, 
which is $\sim 300$ times shorter than the gravitational-wave 
(and pressure-oscillation) period $\tau_{\rm
gw} = 1/f \sim 0.01$ s.  This $\tau_{\rm sound} \ll \tau_{\rm gw}$
means that we can approximate
the oscillations of stress and strain in the test mass, induced by 
the oscillating pressure $P$, as {\it quasistatic}.  It seems reasonable
to expect this approximation to produce a fractional error 
\begin{equation}
\varepsilon_{\rm quasistatic} \alt {\tau_{\rm gw}\over\tau_{\rm sound}} = 
{f_{\rm sound}\over f} \sim {1\over 300}
\label{QuasistaticError}
\end{equation}
in our final answer for the thermoelastic noise $S_q(f)$. Here
\begin{equation} 
f_{\rm sound} = {1\over\tau_{\rm sound}} \sim {c_s\over {\rm min}(a,H)}
\simeq 30,000 {\rm Hz} 
\label{fsound}
\end{equation}
for the currently contemplated LIGO-II test masses: sapphire with $a \sim H
\sim 14$ cm.  

{\it Second:} The timescale
for diffusive heat flow to alter the temperature distribution, 
$\tau_T \sim C_V\rho r_o^2/\kappa \sim 100$ s, is $\sim 10^4$ times longer than
the pressure-oscillation period $\tau_{\rm gw}$ (here $C_V \simeq 7.9 \times
10^6$ erg g$^{-1}$ K$^{-1}$ is the specific
heat per unit mass at constant volume, $\rho \simeq 4.0$ g/cm$^3$ 
is the density, $r_o \sim 4$ cm  is the spot
size and $\kappa\simeq 4.0\times 10^6$ erg cm$^{-1}$ s$^{-1}$ K$^{-1}$ 
is the thermal conductivity, and our values are for a 
sapphire test mass).  This 
$\tau_T \gg \tau_{\rm gw}$ means that, when
computing the oscillating temperature distribution, we can  
approximate the oscillations of stress, strain and temperature as {\it
adiabatic} (negligible diffusive heat flow).  The only place that heat flow
must be considered is in the volume integral (\ref{WdissFD}) for the
dissipation.  The dominant contribution to that volume integral will
come from a region with radius $\sim r_o$ and thickness $\sim r_o$ near
the beam spot.  The region of the integral in which the adiabatic 
approximation breaks down is predominantly a thin ``boundary layer'' near 
the beam spot with radius $r_o$ and thickness of order the distance that
substantial heat can flow in a time $\sim 1/\tau_{\rm gw} = 1/f$, i.e.,
thickness of order
\begin{equation}
r_{\rm heat} = \sqrt{\kappa\over C_V\rho f} \simeq 0.4 
{\rm mm}\sqrt{100 {\rm
Hz}\over f} \quad \hbox{for sapphire.}
\label{rheatflow}
\end{equation}
This region of adiabatic breakdown is a fraction $\sim r_{\rm heat}/r_o$
of the region that contributes substantially to the integral, so we expect
a fractional error 
\begin{equation}
\varepsilon_{\rm adiabatic} \sim {r_{\rm heat}\over r_o} \sim 
0.01
\label{AdiabaticError}
\end{equation}
in $S_q(f)$ due to breakdown of the adiabatic approximation. 
 
The {\it quasistatic} approximation permits
us, at any moment of time $t$, to compute the test mass's internal displacement
field $\vec u$, and most importantly its expansion 
\begin{equation}
\Theta = \vec \nabla \cdot \vec u\;,
\label{Theta}
\end{equation}
from the equations of static stress balance 
(Eq. (7.4) of LL \cite{LL}) 
\begin{equation}
(1-2\sigma) \nabla^2 \vec u + \vec\nabla (\vec\nabla\cdot \vec u) = 0\;
\label{stressbalance}
\end{equation}
(where $\sigma$ is the Poisson ratio), 
with the boundary condition that the normal pressure
on the test-mass face be $P(r,t)$ [Eq.\ (\ref{pressure})] and that all other
non-tangential stresses vanish at the test-mass surface.  Once $\Theta$ has
been computed, we can evaluate the temperature perturbation $\delta T$ from
the law of {\it adiabatic} temperature change
(Eq.\ (6.5) of LL \cite{LL})
\begin{equation}
\delta T = {- \alpha_l E T\over C_V\rho(1-2\sigma)} \Theta \;;
\label{adiabaticT}
\end{equation}
here $\alpha_l$ is the {\it linear} thermal
expansion coefficient, $E$ is Young's modulus and $C_V$ is the 
specific heat
per unit mass at constant volume.\footnote{LL use the 
volumetric thermal expansion
coefficient $\alpha = 3 \alpha_l$ and the specific heat per unit volume $C_v =
\rho C_V$.}  This temperature perturbation can then be plugged into Eq.
(\ref{WdissFD}) to obtain the dissipation $W_{\rm diss}$ as an integral over
the gradient of the expansion
\begin{equation}
W_{\rm diss} = \kappa T \left({E\alpha_l\over(1-2\sigma)C_V\rho}\right)^2
\left\langle \int (\vec\nabla\Theta)^2 \; rd\phi dr dz\right\rangle\;.
\label{WdissTheta}
\end{equation}
This $W_{\rm diss}$ can be inserted into Eq.\ (\ref{SqFD}) to obtain the
thermoelastic noise.

\section{Infinite Test Masses}
\label{sec:InfTM}
\subsection{Dissipation and noise computed via BGV techniques}

We illustrate the above computational procedure by using it to verify
the BGV \cite{BGV} 
result for thermoelastic noise in the case where each test mass is
arbitrarily large compared to the spot size.

Following BGV, we approximate the test mass as an infinite half space.
Then the solution to the quasistatic stress-balance equation
(\ref{stressbalance}) is given by a Green's-function
expression [LL Eq.\ (8.18) with $F_x = F_y = 0$, $F_z = P(r,\phi)$],
integrated over the surface of the test mass.  Taking the divergence of that
expression [or, equivalently, taking the divergence of Eq.\ (39) of BGV], 
we obtain the following equation for the pressure-induced expansion:
\begin{eqnarray} 
&&\Theta = -{(1+\sigma)(1-2\sigma)F_o\over \pi^2 r_o^2 E} \cos(\omega t)
\nonumber\\
&& \times \; z
\int\int_{-\infty}^{+\infty}dx' dy' {e^{-({x'}^2 +
{y'}^2)/r_o^2} \over [(x-x')^2 + (y-y')^2 + z^2]^{3/2}}\;,
\label{Theta1}
\end{eqnarray}
where we have converted from polar coordinates to Cartesian coordinates.
Following a clever procedure implicit in the BGV analysis [in going from their
Eq.\ (39) to (40)], we insert into the integral (\ref{Theta1}) an integral of
the Dirac delta function written as
\begin{eqnarray}
\int_{-\infty}^{+\infty}&& \delta(x-x'-x'')dx'' \nonumber\\
&&= {1\over2\pi}\int\int_{-\infty}^{+\infty}e^{ik_x(x-x'-x'')}dk_x dx''
\label{Dirac}
\end{eqnarray}
and a similar expression for $\int \delta(y-y'-y'')dy''$, 
and we rewrite $x-x'$ and $y-y'$ in
the denominator as $x''$ and $y''$,
thereby obtaining a new version of (\ref{Theta1}) with integrals over $k_x,
k_y, x', y', x'', y''$.  The integrals over $x',y',x'',y''$ are then readily
carried out analytically (they are well-known Fourier transforms), to yield
Eq.\ (40) of BGV:\footnote{Note that our notation differs slightly from that of
BGV: Our $x$ is their $z$, our $z$ is their $x$, and they have factored out
the $\cos\omega t$, which they write as $e^{i\omega t}$.}
\begin{eqnarray} 
\Theta &=& - {(1+\sigma)(1-2\sigma)F_o\over2\pi^2 E}\cos\omega t \nonumber\\
&&\times
\int\int_{-\infty}^{+\infty} e^{-k_\perp^2 r_o^2/4} e^{-k_\perp z} e^{i(k_x x 
+ k_y y)} dk_x dk_y\;,
\label{Theta2}
\end{eqnarray}
where $k_\perp \equiv \sqrt{k_x^2 + k_y^2}$. 

It is straightforward to take the gradient of this expression, square it
(with one term an integral over $k_x,k_y$ and the other over $k'_x,k'_y$),
and integrate over $x$ and $y$ (from $-\infty$ to $+\infty$) and over $z$
(from $0$ to $\infty$); the result is $\int (\vec\nabla\Theta)^2 dx dy dz$
expressed as an integral over $x,y,z,k_x,k_y,k'_x,k'_y$.  The integrals can be
done easily, first over $z$ to get $1/(k_\perp + k_\perp')$, then over $x$ and
$y$ to get Dirac delta functions, then over the $k$'s.  The result, when
inserted into Eq.\ (\ref{WdissTheta}), is
\begin{equation}
W_{\rm diss} = {(1+\sigma)^2 \kappa \alpha_l^2 T\over
\sqrt{2\pi}C_V^2\rho^2r_o^3}F_o^2\;.
\label{WdissFinal}
\end{equation}
By then inserting this into Eq.\ (\ref{SqFD}), we obtain the BGV result for
the thermoelastic noise [their Eq.\ (12)]
\begin{equation}
S_q^{\rm ITM}(f) 
= {8(1+\sigma)^2 \kappa \alpha_l^2 k_{\rm B} 
T^2 \over \sqrt{2\pi}\, C_V^2 \rho^2
r_o^3 \omega^2}\;.
\label{SqBGV}
\end{equation}
Here the superscript ITM means for an ``infinite test mass.''

\subsection{Derivation via BHV techniques}

Equation (\ref{WdissFinal}) for $W_{\rm diss}$ can also be derived
in cylindrical coordinates $(r,z,\phi)$
using the techniques of BHV \cite{bondu}:  The
displacement $\vec u$ has components [BHV Eqs.\ (5) and (6) with the
denominator in (5) corrected from $\mu$ to $\mu+\lambda$ and with
$\beta=\alpha$; see passage following BHV Eq.\ (8)] 
\begin{eqnarray}
u_r &=& \int_0^{\infty} \alpha(k)\left(1-{\lambda+2\mu \over\lambda+\mu} +
kz\right) e^{-kz} J_1(kr) k dk\;,\nonumber\\
u_z &=&  \int_0^{\infty} \alpha(k)\left(1+{\mu \over\lambda+\mu} + 
kz\right) e^{-kz} J_0(kr) k dk\;,\nonumber\\ 
u_\phi &=&0\;, 
\label{uInfCyl}
\end{eqnarray}
where 
\begin{equation}
\alpha(k) = {e^{-k^2 r_o^2/4}\over 4\pi\mu k} F_o \cos\omega t\;
\label{alphak}
\end{equation}
[BHV Eq.\ (11), with the overall sign corrected from $-$ to $+$, with
$w_o=\sqrt2 r_o$ cf.\ BHV Eq.\ (2), and with
$\cos\omega t$ inserted because our method of applying the 
fluctuation-dissipation theorem is dynamical while BHV's method is static].
In Eqs.\ (\ref{uInfCyl})
the $J_n$ are Bessel functions and 
$\lambda$ and $\mu$ are the Lam\'e coefficients (and $\mu$ is also
the shear modulus), which are related to the 
Young's modulus $E$ and the Poisson ratio $\sigma$ by
\begin{equation}
\lambda = {E\sigma\over(1-2\sigma)(1+\sigma)}\;, \quad
\mu = {E\over2(1+\sigma)}\;.
\label{Lame}
\end{equation}  

The divergence of the displacement (\ref{uInfCyl}) is
\begin{equation}
\Theta 
= - {2\mu\over\lambda+\mu}
\int_0^{\infty}
\alpha(k) e^{-kz} J_0(kr) k^2 dk\;.
\label{ThetaInf}
\end{equation}
The nonzero components of the gradient of this expansion are
\begin{mathletters}
\label{gradThetaInfCyl}
\begin{eqnarray}
{\partial\Theta\over\partial r} 
= {2\mu\over\lambda+\mu}
\int_0^{\infty}
\alpha(k) e^{-kz} J_1(kr) k^3 dk\;,\\
{\partial\Theta\over\partial z}
= {2\mu\over\lambda+\mu}
\int_0^{\infty}
\alpha(k) e^{-kz} J_0(kr) k^3 dk\;.
\end{eqnarray}
\end{mathletters}
By squaring the gradient, integrating over the interior of the test mass,
and using the relations
\begin{equation}
\int_0^\infty J_n(k r) J_n(k' r) rdr
= {\delta(k-k')\over k}\;
\label{BesselDelta}
\end{equation}
(which follow from the Fourier-Bessel integral),
and by replacing the Lam\'e coefficients by the Poisson ratio and 
Young's modulus
[Eqs.\ (\ref{Lame})], and inserting the resulting $\int (\vec\nabla\Theta)^2 r
d\phi dr dz$ into expression (\ref{WdissTheta}), we obtain the same
result (\ref{WdissFinal}) as we got using BGV techniques. 
By inserting this into Eq.\ (\ref{WdissTheta}), we obtain the thermoelastic
noise (\ref{SqBGV}).

\section{Finite Sized Test Masses}
\label{sec:FiniteTestMasses}

\subsection{BHV Solution for Displacement}
\label{subsec:BHVDisplacement}

Consider a finite sized, cylindrical test mass with radius $a$ and thickness 
$H$, and with the Gaussian shaped light spot centered on the 
cylinder's circular face.  For this case, Bondu, Hello and Vinet (BHV)
\cite{bondu} have constructed a rather accurate but approximate 
solution of the static elasticity equations. Unfortunately,
their solution 
satisfies the wrong boundary conditions and thus must be
corrected:

The error 
arises when BHV expand the Gaussian-shaped pressure (\ref{pressure})
as a sum over Bessel functions.  BHV incorrectly
omit a uniform-pressure term from the sum.  As a result, the pressure that
they imagine applying to the test-mass face [their Eq.\ (18)],
\begin{equation}
P_{\rm BHV}(r) = F_o \cos\omega t \sum_{m=1}^{\infty} p_m J_0(k_m r)\;
\label{pressureBHV}
\end{equation}
[where $J_0$ is the Bessel function of order zero, $k_m$ is related to the
$m$'th zero $\zeta_m$ of the order-one Bessel function
$J_1(x)$  by $k_m = \zeta_m/a$, and 
$p_m$ are constant coefficients given below], has a vanishing surface
integral
\begin{equation}
\int_0^a P_{\rm BHV} 2\pi r dr = 0\;.
\label{VanishingForce}
\end{equation}
In other words,
their applied pressure (\ref{pressureBHV})
is equal to the desired pressure $P(r)$ [Eq.\ (\ref{pressure})] minus an equal
and opposite net force $F_o \cos(\omega t)$ applied uniformly over the
test-mass face:
\begin{equation}
P_{\rm BHV}(r) = P(r) - p_0 F_o \cos\omega t\;;
\label{pressureRelation}
\end{equation}
\begin{equation}
p_0 \equiv {1 \over \pi a^2} \;.
\label{p0}
\end{equation}
[Recall that we are approximating $1-e^{-a^2/r_0^2}$ by unity; see discussion
following Eq.\ (\ref{qdef}).]

It is evident, then, that to get the correct distribution of 
elastic displacement $\vec u$ inside the test mass, we must add to the
BHV displacement a correction.  This correction is the displacement
caused by the spatially uniform pressure $p_0 F_o\cos\omega t$ on
the test-mass face.  That uniform pressure causes the test mass
to accelerate with acceleration $\vec a =
[(F_o\cos\omega t)/M] \vec e_z$, where $M = \pi a^2 H \rho$ is 
the mass of the test mass and $\rho$ is its density.  
In the reference frame of the accelerating test mass,
all parts of the test mass feel a ``gravitational'' acceleration $g \vec e_z$
equal and opposite to $\vec a$, i.e.\ $g = -(F_o\cos\omega t)/M$, 
(which can be treated as quasistatic, though it oscillates at frequency
$\omega$).  Thus, the displacement is the same as would occur if the
test mass were to reside in the gravitational field $g \vec e_z$ with
a uniform pressure on its face counteracting the force of gravity.
The solution for this displacement is given
by LL \cite{LL} (Problem 1, page 18).\footnote{LL seek to solve a problem
in which (in the presence of the uniform gravitational acceleration), 
instead of having a uniform pressure applied to the face of
the cylindrical test mass, the face has vanishing displacement.  Their
solution actually satisfies our desired boundary conditions but not theirs;
therefore, they comment on it being inaccurate near the test-mass face.
For our problem it is accurate.} 
Translating into our notation and converting from the Young's modulus and
Poisson ratio to the Lam\'e coefficients via Eq.\ (\ref{Lame}), we obtain: 
\begin{mathletters}
\label{DisplacementCorrection}
\begin{eqnarray}
   {\delta u_r\over F_o \cos\omega t} &=& 
\frac{\lambda p_0 r}{2 \mu (3\lambda + 2\mu)}\left( 1-\frac{z}{H} 
\right)\;, \\
   {\delta u_z\over F_o \cos\omega t} &=& 
\frac{\lambda p_0 r^2}{4\mu H (3\lambda+2\mu)}
-\frac{(\lambda+\mu)p_0}{\mu (3\lambda+2\mu)}
\left( z-\frac{z^2}{2H} \right)\;. \nonumber\\
\end{eqnarray}
\end{mathletters}

The total corrected displacement, in cylindrical coordinates, is 
\begin{equation}
u_r = u_r^{\rm BHV} + \delta u_r\;, \quad
u_z = u_z^{\rm BHV} + \delta u_z\;, \quad
u_{\rm \phi} = 0\;,
\label{Displacement}
\end{equation}
where $u_j^{\rm BHV}$ is the BHV displacement  
[their Eqs.\ (15) plus (25) and (17) plus (26)]:
\begin{mathletters}
\label{DisplacementBHV}
\begin{eqnarray}
&&{u_r^{\rm BHV}(r,z)\over F_o\cos\omega t}  =  
\frac{\lambda+2\mu}{2\mu (3\lambda+2\mu)}(c_0r+c_1 rz) \nonumber\\
&&\quad+
\sum_{m=1}^{\infty} A_m(z) J_1(k_m r)\;, \\
&&{u_z^{\rm BHV}(r,z)\over F_o\cos\omega t} = 
-\frac{\lambda}{\mu (3\lambda+2\mu)}\left( 
c_0z+\frac{c_1z^2}{2} \right) \nonumber\\ 
&&\quad-\frac{\lambda+2\mu}{4\mu(3\lambda+2\mu)}c_1 r^2   
+\sum_{m=1}^{\infty} B_m(z) J_0(k_m r)\;,\\
&&u_\phi^{\rm BHV}(r,z) = 0\;.
\end{eqnarray}
\end{mathletters}
Here
the coefficients $c_0$ and $c_1$ are
[equations following Eqs.\ (24) and (26) of BHV]
\begin{equation}
c_0 = 6{a^2\over H^2} \sum_{m=1}^{\infty} {J_0(\zeta_m) p_m \over \zeta_m^2}\;, \quad
c_1 = {-2 c_0\over H}\;,
\label{c0c1}
\end{equation}
and $A_m$ and $B_m$ are the following functions of $z$ 
[Eqs.\ (19) and (20) of BHV]
\begin{eqnarray}
  A_m(z) &=& \gamma_m e^{-k_m z} + \delta_m e^{k_m z} \nonumber\\
&&+ \frac{k_m z}{2}
\frac{\lambda+\mu}{\lambda+2\mu}\left( \alpha_m e^{-k_m z} + \beta_m e^{k_m z} \right)
\label{Am} \\
  B_m(z) &=& \left[ \frac{\lambda+3\mu}{2(\lambda+2\mu)}\beta_m-\delta_m \right] 
e^{k_m z} \nonumber\\
&&+ \left[ \frac{\lambda+3\mu}{2(\lambda+2\mu)}\alpha_m+\gamma_m \right] 
e^{-k_m z} \cr
 & & +\frac{k_m z}{2}\frac{\lambda+\mu}{\lambda+2\mu}\left( \alpha_m e^{-k_m z}
-\beta_m e^{k_m z} \right) \ ,
\end{eqnarray}
where $\alpha_m$, $\beta_m$, $\gamma_m$ and $\delta_m$ are constants given by
[Eqs.\ (21)--(24) of BHV]:
\begin{mathletters}
\label{alphamTodeltam}
\begin{eqnarray}
  Q_m &&= \exp(-2 k_m H) \\
  \alpha_m &&= \frac{p_m (\lambda+2\mu)}{k_m \mu (\lambda+\mu)}
\frac{1-Q_m+2 k_m H Q_m}{(1-Q_m)^2-4k_m^2 H^2 Q_m} \\
  \beta_m &&= \frac{p_m (\lambda+2\mu)Q_m}{k_m \mu (\lambda+\mu)}
\frac{1-Q_m+2 k_m H}{(1-Q_m)^2-4k_m^2 H^2 Q_m} \\
  \gamma_m &&= -\frac{p_m}{2 k_m \mu (\lambda+\mu)} \nonumber\\
&&\times \frac{[2 k_m^2 H^2 (\lambda+\mu)
+ 2\mu k_m H]Q_m+\mu (1-Q_m)}{(1-Q_m)^2-4k_m^2 H^2 Q_m} \\
  \delta_m &=& -\frac{p_m Q_m}{2 k_m \mu (\lambda+\mu)} \nonumber\\
&&\times\frac{2k_m^2 H^2 (\lambda+\mu)
-2\mu k_m H -\mu (1-Q_m)}{(1-Q_m)^2-4k_m^2 H^2 Q_m} \ ,
\end{eqnarray}
\end{mathletters}
with [equation following Eq.\ (18) in BHV]
\begin{equation}
p_m=\frac{2}{a^2 J_0^2(\zeta_m)}\int_0^a {e^{-r^2/r_o^2}\over\pi r_o^2}
\, J_0(k_m r)\, r\,dr \ .
\label{pmDef}
\end{equation}
In the spirit of our approximating $1-e^{-a^2/r_o^2}$ by unity [discussion
following Eq.\ (\ref{qdef})], BHV suggest approximating 
the upper limit of this integral by 
$\infty$; the 
integral can then be done analytically, 
giving [equation preceding Eq.\ (19) of BHV]
\begin{equation}
	p_m=\frac{\exp(-k_m^2 r_o^2/4)}{\pi a^2 J_0^2(\zeta_m)}\;.
\label{pmDefa}
\end{equation}
This is a good approximation to the exact formula (\ref{pmDef}) for small $m$
(which turn out to give the dominant contribution to the noise),
but for large $m$ it can severely underestimate $p_m$. 

\subsection{Expansion and the integral of its squared gradient}

It is straightforward to compute the expansion $\Theta = \vec\nabla \cdot \vec
u$ and the components of its gradient from expressions 
(\ref{DisplacementCorrection}), (\ref{DisplacementBHV}) 
and (\ref{Displacement});
the results are
\begin{eqnarray}
  {\Theta(r,z)\over F_o\cos\omega t} &=& 
-\frac{p_0}{3\lambda+2\mu}\left( 1-\frac{z}{H} \right) +
\frac{2(c_0+c_1z)}{3\lambda+2\mu}\nonumber\\
&&+\sum_m [k_m A_m(z)+B_m'(z)] \,
J_0(k_m r) \;, 
\label{ThetaFinite}
\end{eqnarray}
and
\begin{mathletters}
\label{GradThetaFinite}
\begin{eqnarray}
  \frac{\partial \Theta/\partial r}{F_o\cos\omega t}
&=& -\sum_m k_m [k_m A_m(z)+B_m'(z)] 
J_1(k_m r) \; , \\
  \frac{\partial \Theta/\partial z}{F_o\cos\omega t} 
&=& \frac{2\tilde c_1}{3\lambda+2\mu}+
\sum_m [k_m A_m'(z)+B_m''(z)]\, J_0(k_m r)\;  , \nonumber \\
\end{eqnarray}
\end{mathletters}
where the primes denote derivatives with respect to $z$ and the coefficient
$\tilde c_1$ is
\begin{equation}
\tilde c_1 = c_1 + {p_0\over2H}\;.
\label{tildec1}
\end{equation}

Using the (nonstandard) orthogonality relations 
\begin{eqnarray}
   \int_0^a r\, J_1(k_m r)\, J_1(k_n r)\, dr &=& \frac{a^2}{2}\, J_0^2(\zeta_m) 
\, \delta_{mn}\;, \\
   \int_0^a r\, J_0(k_m r)\, J_0(k_n r)\, dr &=& \frac{a^2}{2}\, J_0^2(\zeta_m) 
\, \delta_{mn}\;, \\
   \int_0^a r\, J_0(k_m r)\, dr &=& 0 \ ,
\end{eqnarray}
the volume integral of $(\vec\nabla\Theta)^2$
can be evaluated analytically. 
The result, after some algebra and after averaging
$\cos^2 \omega t$ to 1/2, is
\begin{eqnarray} 
&&{1\over F_o^2}
\left\langle\int (\vec\nabla\Theta)^2 rd\phi dr dz\right\rangle = 
\frac{2\pi a^2 \tilde{c}_1^2 H}{(3\lambda+2\mu)^2}\nonumber\\
&&\quad+\frac{\pi a^2}{2(\lambda+\mu)^2}
\sum_{m=1}^{\infty} {k_m p_m^2(1-Q_m) J_0^2(\zeta_m) 
\over \left[ (1-Q_m)^2-4H^2 k_m^2 Q_m \right]^2} 
\label{IntGradThetaTwo} \\
&&\quad\quad\times 
\left[(1-Q_m)^2 (1+Q_m)+8H k_m Q_m(1-Q_m)\right. \nonumber\\
&&\quad\quad\quad\left. +4H^2 k_m^2 Q_m(1+Q_m)\right]
 \;.\nonumber
\end{eqnarray}

\subsection{Thermoelastic noise}

Inserting Eq.\ (\ref{IntGradThetaTwo}) into Eq.\ (\ref{WdissTheta}) 
and then into Eq.\ (\ref{SqFD}), and using Eqs.\ (\ref{Lame}) for
the Lam\'e coefficients, we obtain for the spectral density of 
thermoelastic noise in a finite sized test mass:
\begin{equation} 
S_q^{\rm FTM} = C^2_{\rm FTM} S_q^{\rm ITM}\;.
\label{SqFTM}
\end{equation}
Here $S_q^{ITM}$ is the BGV result (\ref{SqBGV}) for the spectral density 
for an infinite test mass, and $C^2_{\rm FTM}$ is the following 
finite-test-mass correction to the 
spectral density:
\begin{eqnarray}
&&C^2_{\rm FTM} = {(2\pi)^{3/2}r_o^3 \over a^3 } \left\{ {a^5 H \tilde
c_1^2\over(1+\sigma)^2} \right.\nonumber\\
&&\quad+ \sum_{m=1}^\infty {a^5  
 k_m p_m^2(1-Q_m) J_0^2(\zeta_m) 
\over \left[ (1-Q_m)^2-4H^2 k_m^2 Q_m \right]^2} 
\label{CFTM} \\
&&\quad\quad \times
\left[(1-Q_m)^2 (1+Q_m)+8H k_m Q_m(1-Q_m)\right.\nonumber\\
&&\quad\quad\quad\left.\left. +4H^2 k_m^2 Q_m(1+Q_m) \right] \right\}. 
\nonumber
\end{eqnarray}

\begin{figure}
\epsfxsize=3.2in\epsfbox{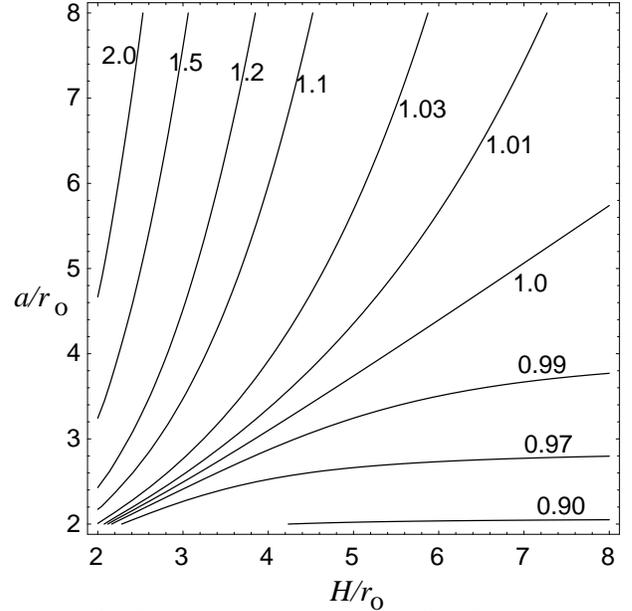}
\caption{Contour plot of the correction $C_{\rm FTM}$
to the thermoelastic amplitude noise 
$\sqrt{S_q(f)}$ due to the finite size of the test mass [Eqs.\ (\ref{SqFTM})
and (\ref{CFTM})]. This correction is shown as a function of the
test-mass radius $a$ and thickness $H$, both measured in units of the 
beam-spot radius $r_o$ (the radius at which the light beam's energy flux has 
dropped to $1/e$ of its central value.)  
\label{FigContour}
}
\end{figure}

The square root, $C_{\rm FTM}$, of this
finite-test-mass correction is plotted in Fig.\ \ref{FigContour} 
as a function of the test-mass thickness $H$ and radius $a$ measured in units
of the beam-spot radius $r_o$.  
[One can easily show from Eq.\ (\ref{CFTM}) 
that $C_{\rm FTM}$ depends on $H$, $a$
and $r_o$ only through the dimensionless ratios $H/r_o$ and $a/r_o$, as must
be the case on dimensional grounds.] 
Notice that
the noise is larger, at fixed $r_o$, for large-$a$, small-$H$
test masses (thin disks) than for small-$a$, large-$H$ test masses
(long cylinders).  However, for plausible parameters the difference is 
only a few tens of per cent.  The reason for the greater noise in
a thin disk is that it experiences greater deformation, when a force
acts at the center of its face, than does a long cylinder, and thus the
integral (\ref{WdissTheta}), to which the noise is proportional, is larger.
(See, e.g., Sec.\ 12 of \cite{LL}, or Sec.\ 305 of \cite{Love}.)

\begin{figure}
\epsfxsize=3.2in\epsfbox{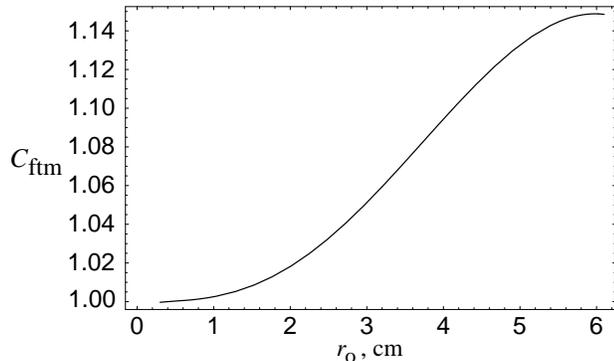}
\caption{Correction $C_{\rm FTM}$ to the thermoelastic amplitude 
noise as a function of the beam-spot radius $r_o$, for test masses with the
parameters currently being contemplated for LIGO-II:
30 kg sapphire cylinders with radius $a=14$ cm and thickness $H= 12.2$ cm. 
\label{Fig1}
}
\end{figure}

The current ``straw-man'' (``reference'') design for LIGO-II includes 
sapphire test masses with $a=14$ cm and $H= 12.2$ cm.  In   
Fig.\ \ref{Fig1} we plot the finite-test-mass correction $C_{\rm FTM}$ as a
function of beam-spot radius $r_o$ (in centimeters) for such test masses
(for which we use the BGV values of the parameters $\alpha = 5.0\times
10^{-6}{\rm K}^{-1}$, $\kappa = 4.0 \times 10^6$ erg K$^{-1}$ cm$^{-1}$ 
s$^{-1}$, $\rho = 4.0$ g/cm$^3$, $C_V = 7.9\times 10^6$ erg g$^{-1}$ K$^{-1}$,
$E= 4\times 10^{12}$ erg/cm$^3$, $\sigma = 0.29$).  Although we continue
our plot up to $r_o = 6$ cm, it may be impractical or undesirable to
operate with $r_o$ much larger than $4$ cm.  Two reasons for this
are: (i) Each time the light beam encounters a test mass, a fraction
$\sim e^{-a^2/r_o^2}$ of its power is lost around the test-mass sides 
(``diffraction
losses''); keeping this below $\sim10$ ppm requires $r_o \alt 4$
cm. (ii) There are practical limitations $R \alt 50$ km on the radii of 
curvature of the test-mass mirrors; if the beam waist is half way between
the mirrors of an arm's optical cavity so the spot sizes $r_o$ are the same 
on the two mirrors, and if $R$ is significantly larger than
the arm length $L=4$ km, then the spot sizes are
$r_o \simeq (\lambda^2 LR/8\pi^2 )^{1/4}$ (where
$\lambda = 1.06\mu$m is the light wavelength), 
so $R\alt 50$ km requires $r_o \alt 4$ cm.

For the plausible range $r_o \alt 4$ cm,
Fig.\ \ref{Fig1} shows that the finite-test-mass correction to the 
amplitude noise
is $\alt 10$ per cent.

\subsection{Errors in our analysis}

There are three significant sources of error in our analysis.  We
expect them to produce a net error in $C_{\rm FTM}$ and thence in
the test-mass noise $\sqrt{S_q^{\rm FTM}}$ that is $\alt 1$
per cent, for the expected LIGO-II parameter regime ($a\sim 14$ cm,
$H\sim 12$ cm, $r_o \alt 4$ cm).  More specifically:

One error source is the quasistatic approximation.  We have already
estimated this as producing a fractional error $\varepsilon_{\rm quasistatic} 
\alt 0.003$ in $S_h$ [Eq.\ (\ref{QuasistaticError})],
and the error in $\sqrt{S_q}$ will be half this, $\alt 0.0015$. 

The second error source is the adiabatic approximation.  We have already
estimated that this produces a fractional error $\varepsilon_{\rm adiabatic} 
\sim 0.01$ in $S_q$ [Eq.\ (\ref{AdiabaticError})], 
and the error in $\sqrt{S_q}$ will be half this, $\alt 0.005$.

The third error source is one that we have not discussed: A failure of
the elastic displacement (\ref{DisplacementBHV}) to satisfy the boundary
condition $T_{rr} = 0$ on the test mass's cylindrical sides, $r=a$.  As
was discussed by BHV \cite{bondu}, 
the $c_0$ and $c_1$ terms in the displacement (\ref{DisplacementBHV})
are a correction to the leading-order displacement, designed to improve
the satisfaction of the $T_{rr}(a) = 0$ boundary condition.  We shall
refer to these terms as the ``Saint-Venant correction'' \cite{bondu}.
In our final answer for 
$S_q(f)$ [Eqs.\ (\ref{SqFTM}) and (\ref{CFTM})], 
this Saint-Venant correction makes a fractional contribution $\alt 6$ per cent,
for LIGO-II type test masses and plausible beam radii $r_o\alt 4$ cm.  
The rms
value of $T_{rr}(a)$ with the Saint-Venant correction included is smaller
than that without the Saint-Venant correction by about a factor 3, so it
is reasonable to expect that the remaining error in $S_q(f)$ due to
$T_{rr}(a)\ne 0$ is $\alt 1/3$
of the Saint-Venant correction, i.e., a remaining fractional error
\begin{equation}
\varepsilon_{\rm SV} \alt {1\over 3} \times 0.06 = 0.02\;.
\label{SaintVenantError}
\end{equation}
The fractional error in $\sqrt{S_q}$ will be half this, $\alt 0.01$ ---
which is larger than the other two errors.

Combining these three errors in quadrature, we expect our formulas for
$\sqrt{S_q^{FTM}}$ to make a net fractional error of magnitude 
\begin{equation}
{\delta C_{\rm FTM}\over C_{\rm FTM}} = 
{\delta \sqrt{S_q^{\rm FTM}} \over \sqrt{S_q^{\rm FTM} }} \alt 0.01\;
\label{neterror}
\end{equation}
for LIGO-II-type test masses and beam-spot radii $r_o \alt 4$ cm.

\section{Conventional Thermal Noise}
\label{sec:Conv}

Because of the boundary-condition error that BHV make 
in solving the elasticity equations (and because of an additional
algebraic error discussed below), their result for the
conventional thermal noise must be corrected.  

The conventional thermal noise is given by Levin's formula
(\ref{SqFD}) with $W_{\rm diss}$ the time-averaged
dissipation produced by
an imaginary part $\Im({E}) = \Phi(\omega) E$ of the Young's
modulus: 
\begin{eqnarray}
W_{\rm diss} &=& 2 \omega \Phi(\omega) \langle U\rangle \nonumber\\
&=& 
\omega \Phi(\omega) \int \left\langle \lambda
\Theta^2 + 2\mu S_{ij}S_{ij}\right\rangle rdrd\phi dz\;.
\label{WdissConv}
\end{eqnarray}
Here $\langle U\rangle$ is the time-averaged elastic energy,
$S_{ij}S_{ij}$ is the square of the strain associated with
the displacement $\vec u$, there is an implied sum over $i$ and $j$, and
the integral is over the test-mass interior;
cf.\ Eq.\ (12) of Ref.\ \cite{levin}.

The expansion $\Theta$ is given by Eq.\ (\ref{ThetaFinite}), and the
components of the strain on the spherical, orthonormal basis $\vec e_r$,
$\vec e_\phi$, $\vec e_z$ are readily computable from the displacement
(\ref{Displacement}), 
(\ref{DisplacementCorrection}), (\ref{DisplacementBHV}) via
[Eqs.\ (A.1)--(A.4) of BHV]
\begin{eqnarray}
S_{rr} &=& u_{r,r}\;, \quad S_{\phi\phi} = {u_r\over r}\;,
\quad S_{zz} = u_{z,z}\;, \nonumber \\
\quad S_{rz} &=& S_{zr} = {1\over2}\left( u_{z,r}+  u_{r,z} \right)\;,
\label{Strain}
\end{eqnarray}
where commas denote partial derivatives. 
By evaluating these strain components, 
inserting them and the expansion (\ref{ThetaFinite}) into Eq.\ 
(\ref{WdissConv}), averaging over time, integrating over the test mass,
and reexpressing the Lam\'e coeffients in terms of the Young's modulus and
Poisson ratio,
we obtain
\begin{equation}
W_{\rm diss} = \omega \Phi(\omega) \left( U_o + \Delta U \right)\;.
\label{WdissConvA}
\end{equation}
Here $U_o$ is given by
\begin{equation}
U_o = {(1-\sigma^2)\pi a^3\over E}\sum_{m=1}^\infty U_m {p_m^2 J_o^2(\zeta_m)
\over\zeta_m}\;,
\label{UoDef} 
\end{equation}
with [equation following Eq.\ (29) of BHV]
\begin{equation}
U_m = {1-Q_m^2 + 4k_m H Q_m \over
(1-Q_m)^2 - 4 k_m^2H^2 Q_m}\;;
\label{UmDef}
\end{equation}
while $\Delta U$ is
\begin{equation}
   \Delta U=\frac{a^2}{6\pi H^3 E} \left[ \pi^2 H^4 p_0^2 + 12\pi H^2
\sigma p_0 s + 72(1-\sigma)s^2 \right]\;,
\label{DeltaUDef}
\end{equation}
with 
\begin{equation}
s=\pi a^2\sum_{m=1}^\infty {p_m J_0(\zeta_m)\over\zeta_m^2}\;.
\label{sDef}
\end{equation}
When the approximation (\ref{pmDefa}) is made for $p_m$, 
$U_o$ takes the form given by BHV [their Eq.\ (30)]
\begin{equation}
U_o = {1-\sigma^2\over \pi a E}\sum_{m=1}^\infty U_m {\exp(-\zeta_m^2
r_o^2/2a^2)\over\zeta_mJ_0(\zeta_m)^2}\;,
\label{UoDefa}
\end{equation}
and $s$ takes the form 
\begin{equation}
   s=\sum_{m=1}^{\infty} \frac{\exp(-\zeta_m^2 r_0^2/4a^2)}{\zeta_m^2 \, 
J_0(\zeta_m)}\;.
\label{sDefa}
\end{equation}
The approximations (\ref{UoDefa}) and (\ref{sDefa})
are rather good for realistic parameter values, despite 
the fact that for
large $m$ Eq.\ (\ref{pmDefa}) is a very poor approximation to $p_m$, because
large $m$ make small contributions to $U_o$ and $s$.

Equations (\ref{DeltaUDef}) and (\ref{sDefa}) for $\Delta U$ differ from
Eq.\ (31) of BHV for two reasons: (i) BHV used the wrong boundary
conditions at the test-mass face [see  beginning of Sec.\ 
\ref{subsec:BHVDisplacement} above]; correcting this
leads to all the terms in Eq.\ (\ref{DeltaUDef}) involving $p_o$.  (ii) BHV
seem to have made an algebraic error: Eqs.\ (\ref{DeltaUDef}) and (\ref{sDefa})
should agree with BHV Eq.\ (31) when $p_o$ is set to zero, but they do not; 
it might be that BHV accidentally omitted the $S_{\phi\phi}^2$ term or the
$S_{rr}^2$ term when evaluating Eq.\ (\ref{WdissConv}).

Inserting Eq.\ (\ref{WdissConv}) into Eq.\ (\ref{SqFD}), we obtain the
BHV expression for the conventional thermal noise [their equation following
Eq.\ (31)]
\begin{equation}
S_q^{\rm FTM}(f) = {8k_{\rm B} T\over\omega} \Phi(\omega) (U_o + \Delta U)\;,
\label{SqConv}
\end{equation}
where (to reiterate) $U_o$ is given by Eqs.\ (\ref{UoDef}) [or (\ref{UoDefa})]
and (\ref{UmDef}), 
while $\Delta U$ is given by Eqs.\ (\ref{DeltaUDef}) and
(\ref{sDef}) [or (\ref{sDefa})]. 

\begin{figure}
\epsfxsize=3.2in\epsfbox{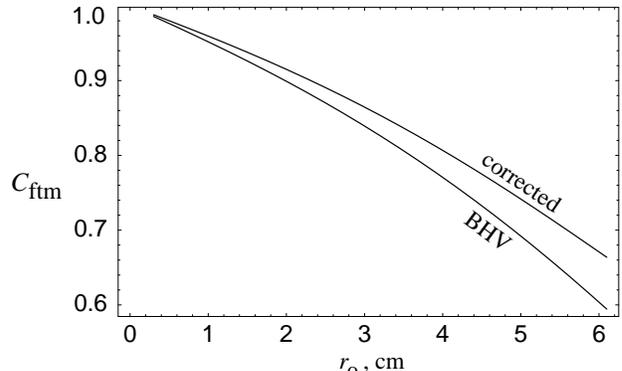}
\caption{$C_{\rm FTM} = \sqrt{S_q^{\rm FTM}}/\sqrt{S_q^{\rm ITM}}$,
the finite-test-mass correction 
to the conventional, homogenous thermal 
noise, as a function of the beam-spot radius $r_o$, for test masses with the
parameters currently being contemplated for LIGO-II:
30 kg sapphire cylinders with radius $a=14$ cm and thickness $H= 12.2$ cm. 
The curve labeled ``BHV'' is the result derived in Ref.\ 
\protect\cite{bondu}
[their Eqs.\ (29), (28), and equation following (31)]; the curve labeled 
``corrected'' is our corrected version of their result [our Eqs.\
(\ref{SqConv}), (\ref{UoDef}) and (\ref{DeltaUDef})].  
\label{FigConv}
}
\end{figure}

If the test mass is infinite in size, then the conventional thermal
noise has the following
form, derived by BHV [their Eq.\ (14) with $w_o = \sqrt{2}\,r_o$, 
which differs from the formula
derived earlier by Levin \cite{levin}---his Eq.\ (2)]:
\begin{equation}
S_q^{\rm ITM} = {4 k_{\rm B} T\over\omega}{1-\sigma^2\over\sqrt{2\pi} E
r_o}\Phi(\omega)\;.
\label{SqConvITM}
\end{equation}
As for thermoelastic noise, we define a finite-test-mass
correction $C^2_{\rm FTM}$ to be the ratio of the finite-test-mass 
spectral density (\ref{SqConv}) to that (\ref{SqConvITM})
for the infinite test mass:
\begin{equation}
C^2_{\rm FTM} = {S_q^{\rm FTM}\over S_q^{\rm ITM}}
\label{CFTMConv}
\end{equation}
We plot the square root of this correction (i.e., the amplitude-noise
correction) as a function of beam-spot radius $r_o$ in Fig.\ \ref{FigConv}
for a LIGO-II type test mass.  We show there two curves, $C_{\rm FTM}$ as
given by the BHV formulae, and as given by our corrected formulae.
Note that the BHV errors have only a small influence: their noise was 
too low by a factor $\alt 5$ per cent when $r_o \alt 4$ cm.

\section{Conclusion}
\label{sec:Conclusion} 

In this paper we have sketched a fairly simple method of analyzing
thermoelastic thermal noise in interferometric detectors, we have used that
method to derive formulas for the noise in cylindrical test masses with
finite radius, thickness, and beam spots, and we have corrected the
corresponding finite test-mass formulas for conventional thermal
noise.  Our formulas should be useful in optimizing the test-mass
designs for interferometric gravitational wave detectors.  

Because thermoelastic noise arises from physical processes associated with
ordinary thermal fluctuations, thermal conductivity and thermal expansion,
and is {\it not} influenced by ``dirty'' processes such as lattice defects
and impurities (except through the easily measured 
conductivity and expansion),
the predictions for thermoelastic noise should be very reliable.  Nevertheless,
experimental tests of the theory would be useful and are being planned.  

Other forms of thermal noise do rely in crucial, ill-understood ways
on dirty processes and thus are far less reliably understood than
thermoelastic noise.  This is especially the case of thermal noise associated
with (inhomogeneous) dissipation in and just beneath the test mass's
dielectric-mirror coatings [for which Levin \cite{levin} predicts,
in the infinite-test-mass limit, 
a dependence $S_q \propto 1/r_o^2$ on beam-spot radius, compared to
$S_q \propto 1/r_o^3$ for thermoelastic noise and $S_q \propto 1/r_o$ for 
conventional, homogeneous thermal noise].  Detailed experimental studies
of these other forms of thermal noise are much
needed as part of the R\&D for interferometric gravitational-wave detectors,
and are being planned.

In some of the planned 
experiments, very small beam radii $r_o$ and/or high frequencies $f$
may be used.  For 
\begin{eqnarray}
r_o \alt r_o^{\rm heat} &\equiv& \sqrt{\kappa\over C_V\rho f}\nonumber\\
&&\simeq 0.4 {\rm mm} \sqrt{100 {\rm Hz}\over f} \quad \hbox{for sapphire,}
\label{road}
\end{eqnarray}
the adiabatic approximation breaks down seriously [cf.\ Eq.\
(\ref{AdiabaticError}) and associated discussion] 
and our analysis of thermoelastic noise must be redone
taking account of the diffusive redistribution of temperature during 
the elastic oscillations.  Some foundations for doing this have been
laid by BGV \cite{BGV}.  For frequencies
\begin{eqnarray}
f\agt f_{\rm sound} &\equiv& {c_s\over{{\rm min}(a,H)}}\nonumber\\
&&\simeq
10^4 {\rm Hz}{10{\rm cm} \over {\rm min}(a,H)} \quad \hbox{for sapphire}\;
\label{fqs}
\end{eqnarray}
(where $c_s$ is the sound speed),
the quasistatic approximation breaks down seriously
[cf.\ Eq.\ (\ref{QuasistaticError})
and associated discussion], and our analysis must be redone taking account
of the finite propagation speed of the test mass's elastic deformations.  

After completing our analysis of thermoelastic noise in finite sized test
masses, we learned that 
Sergey Vyatchanin \cite{vyatchanin} has been carrying out
an analysis of this same
issue, but using somewhat different techniques.  In writing the final version
of this paper, we have benefitted from email exchanges with him.

\section*{Acknowledgments}

For helpful advice we thank Michael Gorodetsky, 
Eric Gustafson, James Hough, David Shoemaker, Jean-Yves Vinet,
Rainer Weiss, and especially Vladimir Braginsky and Sergey Vyatchanin.  
A lively interchange
of email with Vyatchanin has helped us understand more deeply
the issues underlying thermoelastic noise, and we expect that his
manuscript \cite{vyatchanin} will shed valuable new light on those issues.  
This research was supported in part by National Science Foundation Grant
PHY--9900776.


\begin{references}

\bibitem{BGV} V.\ B.\ Braginsky, M.\ L.\ Gorodetsky, and S.\ P.\ Vyatchanin,
Phys.\ Lett.\ A {\bf 264}, 1 (1999); cond-mat/9912139; cited in text as BGV.

\bibitem{ligo} http://www.ligo.caltech.edu/.

\bibitem{bondu} F.\ Bondu, P.\ Hello, and J.-Y.\ Vinet, Phys.\ Lett.\ A {\bf
246}, 227 (1998); cited in text as BHV.  This paper contains a number
of errors, mostly typographical. 
Because the paper is so fundamental to thermal-noise modeling, the following
list of errors may be of help to other researchers.  
(Vinet, private communication, agrees with this list). 
In Eqs.\ (3) and (5)
the denominator $\mu$ should be $\lambda+\mu$ [same as in Eqs. (4) and (6)];
assuming the Landau-Lifshitz sign convention for the stress tensor as in
Appendix A of BHV, the second
boundary condition after Eq.\ (8) should be $\Theta_{zz}(r,z=0) = - p(r)$ and
the last boundary condition after Eq.\ (17) should be $\Theta_{zz}(r,z=0) = -
p(r)$; the overall minus signs in Eqs. (9)--(11) should all be changed to $+$;
there should be a factor $\mu$ in the denominator of Eq.\ (12);
in the equation for
$p_m$, preceding Eq.\ (19), the $\sigma$ in the denominator should be $\pi$;
in Eq.\ (19), in the third term on the right side, $k_m a$ should be $k_m z$
[cf.\ our Eq.\ (\ref{Am})];
Eqs.\ (21)--(24) should take the forms given in our Eqs.\ 
(\ref{alphamTodeltam}); and Eqs. (28) and (31) for $\Delta U$ should
take the forms given in our Eqs.\ (\ref{DeltaUDef}), (\ref{sDef})
and (\ref{sDefa}). 


\bibitem{levin} Yu.\ Levin, Phys.\ Rev.\ D {\bf 57}, 659 (1998).  Note that
Levin made an error in calculating the oscillating
elastic energy, when applying
his method to conventional, homogeneous thermal noise. His Eqs.\ (A5)
and (14) should actually have the form derived
by BHV \cite{bondu} [their Eq.\ (13)], and correspondingly, his   
final formula (1), (15) for the conventional thermal noise should actually
be that derived by BHV [their Eq.\ (14) with $w_o = \sqrt{2}\,r_o$; 
our Eq.\ (\ref{SqConvITM})]. 

\bibitem{callenwelton} H.\ B.\ Callen and T.\ A.\ Welton, Phys.\ Rev.\ {\bf 
83}, 34 (1951). 

\bibitem{LL} L.\ D.\ Landau and E.\ M.\ Lifshitz, {\it Theory of Elasticity},
third edition (Pergamon, Oxford, 1986).

\bibitem{Love} A.\ E.\ H.\ Love, {\it A Treatise on the Mathematical Theory of
Elasticity}, Fourth Edition (Cambridge University Press, Cambridge 1927; 
reprinted by Dover, New York, 1944).

\bibitem{vyatchanin} S.\ P.\ Vyatchanin, paper in preparation.

\end{references}
\end{document}